\documentclass[pra,twocolumn,superscriptaddress,amsmath,amssymb,floatfi,showpacs]{revtex4}

\usepackage{graphicx}

\begin{document}
\title{\textbf{Qubit Coherent Control with Squeezed Light Fields}}

\author{Ephraim Shahmoon$^1$, Shimon Levit}

\affiliation {Department of Condensed Matter Physics, Weizmann
Institute of Science, Rehovot, 76100, Israel}
\author{Roee Ozeri}
\affiliation {Department of Physics of Complex Systems, Weizmann
Institute of Science, Rehovot, 76100, Israel}
\date{\today}

\begin{abstract}
We study the use of squeezed light for qubit coherent control and
compare it with the coherent state control field case. We calculate
the entanglement between a short pulse of resonant squeezed light
and a two-level atom in free space and the resulting operation
error. We find that the squeezing phase, the phase of the light
field and the atomic superposition phase, all determine whether
atom-pulse mode entanglement and the gate error are enhanced or
suppressed. However, when averaged over all possible qubit initial
states, the gate error would not decrease by a practicably useful
amount and would in fact increase in most cases. We discuss the
possibility of measuring the increased gate error as a signature of
the enhancement of entanglement by squeezing.

\end{abstract}

 \maketitle

\section{Introduction}\label{section1}
The coherent control of a two-level atom in free space with resonant
laser fields is important both from a fundamental as well as a
practical standpoint. In particular, for quantum information
processing purposes, where the two-level atom encodes a single
quantum bit (qubit), coherent control accuracy plays a crucial role.
Fault-tolerant quantum error correction schemes, necessary for large
scale quantum computing, demand that the error in a single quantum
gate is below a certain, typically very small, threshold \cite{Knill
2005}.

In recent years the fundamental limitation to qubit control accuracy
imposed by the quantum nature of the control light fields has been
the focus of several studies \cite{VE-K, banacloche, SD-PRA68,
itano, SD-PRA69, Nha-Carmichael}. For a coherent state of light
(CL), where both phase and amplitude fluctuations are at the
standard quantum limit (SQL), it was found that the gate error is
determined by the qubit spontaneous decay rate. At a time shorter
than the qubit decay time, part of that error can be interpreted as
due to entanglement between the control field mode and the qubit
(forward scattering).

This entanglement is a measure of the information on the atomic
state carried away by the pulse field and could, in principle, be
exploited as a resource in quantum communication schemes. For a CL
pulse of duration $\tau$, the entanglement, measured by the tangle
 $T$, was found to be \cite{SD-PRA69}, $T\sim\tau\sim 1/\sqrt{\bar{n}}$,
 where $\bar{n}$ is the average number of photons in the pulse. Thus, in agreement with the
correspondence principle, quantum entanglement becomes very small as the pulse
approaches the classical limit, i.e. $\bar{n}\gg 1$.

It is therefore natural to ask whether a control with non coherent
light field states would reduce or increase both the operation error
and atom-light mode entanglement. In this paper we investigate the
fundamental quantum limitation to the fidelity of qubit coherent
control with squeezed light (SL) fields. As opposed to CL, SL has
phase-dependent quantum fluctuations, where one field quadrature has
reduced and the other increased fluctuations compared with the SQL.
Reduced phase fluctuations of squeezed light have been shown to, in
principle, improve the fundamental limitation to phase metrology
accuracy \cite{Caves81, Grangier87}. The use of squeezed light
states was also shown to be beneficial for continuous variable
quantum key distribution \cite{Ralph,Hillery,Preskill}. Here, we
explore the possible advantage of the use of squeezed light for
qubit coherent control. Both the error of a single-qubit quantum
gate and the entanglement between the atomic qubit and the SL mode
are calculated.

Atom-SL interactions were studied extensively in the past
\cite{DF-rev, DF-book}. In most cases the atomic dynamics was found
to be modified in a squeezing-phase-dependent way. Gardiner
\cite{gar86} found that a two-level atom damped by a squeezed vacuum
reservoir exhibits squeezing-phase-dependent suppression or
enhancement of atomic coherence decay. Carmichael \cite{car87} added
a coherent field that drives the atom and showed that the dynamics
of the atomic coherence depends on the squeezed vacuum and the
coherent driving field relative phase. Both assumed a broadband
squeezed vacuum reservoir whereas the finite bandwidth case was
studied in \cite{ParGar}. Milburn, on the other hand, studied the
interaction of a single-mode SL with an atom using the
Janes-Cummings model and found a squeezing-phase-dependent increase
or decrease of the collapse time \cite{Milburn_JC}.

Here we assume a qubit that is encoded in an atomic, two-level,
superposition and a control field that is realized with a pulse of
SL in a single electromagnetic (EM) field mode and of duration
shorter than the typical atomic decay time, in free space. We find
that the squeezing phase, the phase of the light field and the
atomic superposition phase, all determine whether atom-pulse mode
entanglement and the gate error are enhanced or suppressed. It is
found that, although reduced for certain qubit initial states, when
averaged over all possible states the minimal gate error is
comparable to that with CL fields.

The paper is outlined as follows. Section \ref{section2} describes a
general model of the interaction between a light pulse and an atom
in free space within the paraxial approximation. In Sec.
\ref{section3} the pulse mode is assumed to be in a squeezed state
and its effect on the atom is examined. Atom-pulse entanglement as
well as the operation error are calculated. Some aspects of possible
experimental realization are discussed in Sec. \ref{section4} where
our results are also briefly summarized.

\section{General Model and Approach} \label{section2}
We describe the interaction of a two-level atom with a quantized
pulse of light propagating in free space. Classically the
propagation of such a pulse is well approximated by a paraxial wave
equation. A paraxial formalism for quantized pulses was described in
 \cite{woerdman, SD-PRA68, Deutsch}. We briefly describe the paraxial model following a similar approach to that in
\cite{SD-PRA68}. The derivation of this approximation is outlined in
the appendix, where we start from the basic atom-field Hamiltonian
in free space and get to the model described below.

\subsection{The pulse mode}
Consider a laser pulse traveling in the $z$ direction and
interacting with a well-localized atom or ion at position
$\vec{r}_a$. This pulse is a wave packet of Fourier modes around
some wave vector $\vec{k}_0=k_0\vec{z}$, where $\vec{z}$ is a unit
vector pointing to the $z$ direction, i.e. it is composed of a
carrier wave vector $\vec{k}_0$ and a slowly varying envelope. The
pulse is assumed to propagate with a small diffraction angle, so
that the paraxial approximation is valid. A natural description for
such a pulse is as a superposition of Gaussian beams with a
well-defined transverse area $A$ at the position of the atom, each
with wave number $k_0+k$, where $k\ll k_0$, analogous to a
narrowband wave packet of Fourier modes in 1d. In this approximation
the pulse is effectively a one-dimensional beam with transverse area
$A$ similar to a Gaussian beam. The envelope of the pulse mode,
$\varphi_{f_0}(z)$, is a function of width $c\tau$ where $\tau$ is
the pulse duration and $c$ the speed of light. The field operator
assigned to this pulse mode is $\hat{a}_{f_0}$.

\subsection{Interaction picture and the atom-pulse bipartite system}
We start with the full atom-field Hamiltonian in free space,
\begin{eqnarray}
&&\hat{H}=\hat{H}_F+\hat{H}_A+\hat{H}_{AF} \nonumber\\
&&\hat{H}_A+\hat{H}_F=\frac{1}{2}\
\hbar\omega_{a}\hat{\sigma}_{z}+\sum_{\vec{k} \mu} \hbar\omega_k
\hat{a}^\dagger_{\vec{k} \mu} \hat{a}_{\vec{k} \mu}
\nonumber \\
&&\hat{H}_{AF}=\sum_{\vec{k} \mu}i\hbar \left ( g_{\vec{k} \mu}
\hat{a}_{\vec{k} \mu} - h.c. \right) \left ( \hat{\sigma}_{+} +
\hat{\sigma}_{-} \right ), \label{1}
\end{eqnarray}
where $\vec{k}\mu$ are the indices of a Fourier mode and its
polarization respectively, and $g_{\vec{k} \mu}$ are the atom-mode
couplings. Here the $\hat{\sigma}$ operators are the spin $1/2$
Pauli matrices and $\hbar\omega_a$ is the energy separation between
the two atomic levels. Moving to the interaction picture with
respect to the interaction $\hat{H}_{AF}$ in the rotating wave
approximation, assuming no detuning $\omega_0=k_0c=\omega_a$, and
using the paraxial approximation, we get the Hamiltonian in the
interaction picture (see appendix)
\begin{eqnarray}
\hat{H}_I&=&\hat{H}_S+\hat{H}_{SR} \nonumber\\
\hat{H}_S&=&i \hbar g_{f_0}(t) \left(\ \hat{a}_{f_0}
\hat{\sigma}_{+} - \hat{a}_{f_0}^{\dag}\hat{\sigma}_{-} \right) \nonumber \\
 \hat{H}_{SR}&=&i \hbar \sum_r \left ( g_r \hat{a}_r\hat{\sigma}_{+} - g_r^{\ast}\hat{a}_r^{\dagger}\hat{\sigma}_{-}\right),
\label{H_I}
\end{eqnarray}
where the indices $\{r\}$ denote all the free space EM modes apart
from the $f_0$ pulse mode, with couplings $g_r$ which are in general
time dependent. The $f_0$ mode coupling is
\begin{equation}
g_{f_0}(t)=d\sqrt{\frac{\omega_{0}}{2\epsilon_{0}\hbar A}}
e^{ik_{0}z_a}\varphi_{f_0}(z_a-ct),
 \label{g_f0}
\end{equation}
where $z_a$ is the atom's position. In Eq. (\ref{H_I}) we assume
both the envelope of the pulse $\varphi_{f_0}(z)$ and the dipole
matrix element $d$ to be real and set $z_a=0$. A similar model
Hamiltonian was also used in
 \cite{SD-PRA69, VE-K} where the entanglement between a CL pulse and an atom was
  calculated. Note that the time dependence of the atom-pulse coupling $g_{f_0}(t)$ is
  identical to that of the pulse, the pulse therefore switches the interaction on and off.\\

The relevant \emph{bipartite system} is composed of the atom and the
$f_0$ light mode, coupled by $\hat{H}_S$. The rest of the EM modes
except for $f_0$, $\{r\}$, make up a \emph{reservoir} that interacts
with the bipartite system via the coupling with the atom.

\section{Squeezed Light Pulse operating on an atom} \label{section3}
We consider a pulse of squeezed light operating on an atom. The
pulse's average field changes the atomic superposition coherently.
In a Bloch sphere representation this corresponds to a rotation of
the Bloch vector. The quantum fluctuations of the field lead to
atom-pulse entanglement and add a random component to the average
operation. We wish to analyze and calculate the error caused by
quantum fluctuations as well as the degree of atom-pulse mode
entanglement. The model outlined in Sec. II is sufficiently general
to describe a pulse in any quantum state and its interaction with
the atom. In \cite{SD-PRA69,VE-K} calculations for CL state were
performed. In this Section we will discuss quantum pulses of SL
interacting with a two-level atom. The general formalism is
presented in Sec. III A. In Sec. III B we consider a rotation of the
Bloch vector by 180 degrees about an axis lying in the Bloch sphere
equatorial plane (a $\pi$ pulse) as an example and neglect the
reservoir (i.e. the rest of the EM modes). This allows to obtain
analytic expressions for the atom-pulse entanglement which, in this
case, is the sole generator of gate error. For longer pulses,
reservoir effects become increasingly dominant for both entanglement
and error generation. In Sec. III C we obtain the results in the
presence of the reservoir, i.e. including spontaneous emission into
all modes.

\subsection{Unitary transformations and master equation }
In order to determine the final state of the bipartite photon
pulse-atom system in the presence of the reservoir of free EM modes,
an appropriate master equation should be used.

\subsubsection{Initial state}
Here we assume the light mode $f_0$ to be a single pulse mode in a
squeezed state and choose the following initial state for the EM
field
\begin{equation}
|\alpha,\varepsilon\rangle_{f_0}\otimes|0\rangle_{\{r\}}.\label{9}
\end{equation}
Here
\begin{equation}
|\alpha,\varepsilon\rangle_{f_0}=\hat{D}(\alpha)\hat{S}(\varepsilon)|0\rangle_{f_{0}},
\label{10}
\end{equation}
where
\begin{equation}
\hat{S}(\varepsilon)\equiv
\hat{S}_{f_{0}}(\varepsilon)=e^{\frac{1}{2}(\varepsilon^*
\hat{a}_{f_{0}}^2-\varepsilon \hat{a}_{f_0}^{\dag 2})} \label{11}
\end{equation}
is the squeezing operator with the squeezing parameter
$\varepsilon=re^{i2\phi}$ and
\begin{equation}
\hat{D}(\alpha)\equiv \hat{D}_{f_0}(\alpha)=e^{\alpha
\hat{a}^{\dagger}_{f_0}-\alpha^{\ast} \hat{a}_{f_0}} \label{12}
\end{equation}
is the displacement operator with the field average $\alpha=|\alpha|e^{i\theta}$.

The phases $\phi$ and $\theta$ together with the phases associated
with the initial state of the atomic qubit will play an  important
role in the following.

\subsubsection{Unitary transformations}
It is convenient to transform the problem into a representation
where the initial state of the pulse mode is the vacuum state. To
achieve this, we first use the Mollow transformation \cite{Mollow,
CCT} and move the average field $\alpha$ from the initial state into
the Hamiltonian. This affects only the term $\hat{H}_S$ in the
Hamiltonian (\ref{H_I}) which becomes
\begin{equation}
\hat{H}_S=i \hbar g_{f_0}(t) \left [ \hat{a}_{f_0} \hat{\sigma}_{+}
- h.c. \right]+i \hbar g_{f_0}(t) \left [\alpha \hat{\sigma}_{+} -
h.c. \right]. \label{H_S1}
\end{equation}
We then make a time-independent Bogoliubov transformation,
$\hat{S}^{\dag}(\varepsilon)\hat{H} \hat{S}(\varepsilon)$. This
again affects only $\hat{H}_S$ which becomes,
\begin{eqnarray} \label{hamiltonian1}
 \hat{H}_{S}&=& i\hbar g_{f_0}(t) \left [\hat{\sigma}_{+}\left (c \hat{a}-e^{-2i\phi}s \hat{a}^{\dag} \right ) - h.c. \right]  \nonumber \\
  &+&i \hbar g_{f_0}(t)\left[ \alpha \hat{\sigma}_{+}-\alpha^{\ast}\hat{\sigma}_{-}
  \right].
  \label{H_S}
\end{eqnarray}
Here $\hat{a}\equiv \hat{a}_{f_0}$, $s=\sinh(r)$ and $c=\cosh(r)$.
In this representation the initial state of all the EM modes is the
vacuum state.

The Hamiltonian (\ref{H_S}) includes two types of driving terms. The
second term in (\ref{H_S}) describes an atom driven by  the average
field and therefore contains no EM field operators. The first term
in (\ref{H_S}) describes interaction with the EM squeezed vacuum.
Unlike the coherent state case, it includes terms like
$\hat{a}^{\dag}\hat{\sigma}_{+}$ similar to those that are typically
neglected in the rotating wave approximation. Here however, such
terms appear as a result of squeezing, and indeed vanish for $r=0$.
These terms reflect the fact that unlike the ordinary vacuum the
squeezed vacuum contains photons that can be absorbed by the atom.

\subsubsection{Master equation}
We now examine a particular case of a $\pi$ pulse of duration
$\tau$. In a Bloch sphere representation and within our notations,
the average field of such a pulse rotates the Bloch vector by 180
degrees about an axis, lying in the sphere equatorial plane, in an
angle $\theta$ from the $y$ axis. For a constant interaction
coefficient $g$ this requires $|\alpha| g \tau=\pi/2$. The
time-dependent interaction coefficient $g_{f_0}(t)$ from Eq.
(\ref{g_f0}) includes the pulse profile, which is assumed to be
approximately constant during the time interval $\tau$ and vanishing
elsewhere. We set $t=0$ to be the time at which the pulse reaches
the atom so at $t=\tau$ it has already passed it,
\begin{eqnarray}
&&g_{f_0}(0<t<\tau)\equiv g \equiv \sqrt{\kappa/ \tau} \nonumber \\
&&g_{f_0}(t<0,t>\tau)=0. \label{g(t)}
\end{eqnarray}
Here $\kappa=d^2\omega_0/(2\varepsilon_0\hbar A c)$ is the atomic
decay rate into all the paraxial modes of the same transverse
profile as the $f_0$ pulse mode \cite{SD-PRA68}. The dynamics during
the pulse is then approximately governed by,
\begin{eqnarray}
\hat{H}&=&\hat{H}_S+\hat{H}_{SR} \nonumber \\
 \hat{H}_{S}&=& i \hbar \sqrt{\frac{\kappa}{\tau}}c \hat{W}+ i \hbar \frac{\pi}{2\tau} \left( e^{i\theta}
 \hat{\sigma}_{+}-e^{-i\theta}\hat{\sigma}_{-} \right)\nonumber \\
 \hat{H}_{SR}&=&i \hbar \sum_r \left ( g_r \hat{a}_r\hat{\sigma}_{+} - g_r^{\ast}\hat{a}_r^{\dagger}\hat{\sigma}_{-}\right),
\label{H}
\end{eqnarray}
where $\hat{W}$ is the dimensionless operator that describes the
quantum field part of the atom-pulse interaction,
\begin{equation}
\hat{W}= \left [\hat{\sigma}_{+}\left
(\hat{a}-e^{-2i\phi}\frac{s}{c} \hat{a}^{\dag} \right ) -
 \hat{\sigma}_{-}\left ( \hat{a}^{\dag}-e^{2i\phi}\frac{s}{c} \hat{a} \right ) \right ].
 \label{F}
\end{equation}
The corresponding master equation for the bipartite system density
matrix, $\hat{\rho}(t)$, in the presence of the reservoir is
 \begin{equation}
\frac{d \hat{\rho}}{dt}=-
\frac{i}{\hbar}[\hat{H}_S,\hat{\rho}]-\frac{\Gamma}{2}
\{\hat{\sigma}_{+} \hat{\sigma}_{-}, \hat{\rho}\}+ \Gamma
\hat{\sigma}_{-} \hat{\rho} \hat{\sigma}_{+}, \label{master}
\end{equation}
where $\Gamma$ is the atomic decay rate in free space.

This master equation assumes a Markovian reservoir, i.e. a reservoir
with a very short correlation time. Such assumption is not obvious
for the free space EM reservoir that lacks a single pulse mode, like
the one we consider. This rather artificial reservoir has a
correlation time of the order of the pulse duration $\tau$. In
section IV B of \cite{SD-PRA69}, a similar Markovian equation was
written for a CL pulse and was then solved perturbatively for
$\kappa\tau\ll 1$. In fact, in \cite{thesis} it was found that the
assumption $(g\tau)^2=\kappa\tau\ll 1$ is actually a necessary
\emph{condition} for the validity of the Markovian assumption and
the Master equation (\ref{master}), together with the condition
$\kappa\ll\Gamma$. An intuitive argument for the requirement
$(g\tau)^2=\kappa\tau\ll 1$ can be based on the time scales related
to the pulse mode that is excluded from the reservoir. While the
correlation time of the reservoir is $\tau$, this correlation's
effect on the dynamics should be of the order of $\kappa$. Therefore
Markovian dynamics is obtained by requiring $\kappa\tau\ll 1$.
Markovian time evolution guarantees that the probability of
successive reabsorptions and reemissions of photons from and into
the $f_0$ mode, typical of the Janes-Cummings dynamics, are
negligible as it should be in free space.

The initial bipartite state,
$\hat{\rho}(0)=|\sigma,0\rangle\langle\sigma,0|$, where
$|\sigma\rangle$ denotes any atomic qubit state and $f_0$ is in the
vacuum $0$, is pure and separable. Solving Eq. (\ref{master}) for
$t=\tau$ yields the bipartite density operator $\hat{\rho}(\tau)$
from which properties like atom-pulse entanglement can be
calculated.

\subsection{Entanglement and error probability calculations neglecting the reservoir}

We start by considering a pulse duration which is much shorter than
the atomic decay time $\Gamma^{-1}$. An analytic solution to Eq.
(\ref{master}) can be found in this case by neglecting the
reservoir. In this short pulse limit, the probability to absorb a
photon from the pulse and then to emit one to the reservoir is
negligible. Then, one can indeed neglect any modification to the
atom-pulse entanglement due to the presence of the reservoir.
However, gate error calculated in this approximation is induced only
by the quantum fluctuations of the pulse mode. We note that at least
for the CL case, this error is a relatively small part of the total
error which is mostly due to spontaneous photon emission to the
reservoir EM modes.

\subsubsection{Bipartite wave function calculation}
Neglecting the reservoir, Eq. (\ref{master}) becomes a
Schr\"{o}dinger equation, the solution of which at $t=\tau$ is $
|\psi(\tau)\rangle=\hat{U}(\tau) |\sigma,0\rangle$, where
\begin{equation}
 \hat{U}(\tau)=e^{-\frac{i}{\hbar} \hat{H}_{S}\tau}=e^{\frac{\pi}{2}(\hat{\sigma}_{+}-\hat{\sigma}_{-})+\lambda \hat{W}}
\label{U}
\end{equation}
is the propagator for $t=\tau$. Here $\theta=0$ is chosen and
$\lambda \equiv\cosh(r)\sqrt{\kappa\tau}$. We now take the
assumption $\kappa\tau\ll1$ one step further and require
$\lambda\ll1$. This puts a limit on the magnitude of the squeezing
parameter $r$. As will be shown later, in the paraxial approximation
where $\kappa$ is small and with presently available squeezing
technology, this is a very good approximation. Looking at the
propagator in Eq. (\ref{U}) one notes that initially the matrix
elements of $\hat{W}$ are $O(1)$ so $\hat{U}(\tau)$ can be expanded
in powers of $\lambda$
\begin{equation}
 \hat{U}(\tau)= \hat{U}^{(0)}+\lambda
 \hat{U}^{(1)}+\lambda^2\hat{U}^{(2)}+...,
\label{19}
\end{equation}
where
\begin{eqnarray}
 \hat{U}^{(0)}&=&\hat{U}_0(\xi=1) \nonumber\\
 \hat{U}^{(1)}&=&\int_{0}^{1}d\xi \hat{U}_0(1-\xi)\hat{W}\hat{U}_0(\xi) \label{evol}\\
 \hat{U}^{(2)}&=& \int_{0}^{1}d\xi_2 \int_{\xi_2}^{1}d\xi_1
 \hat{U}_0(1-\xi_1)
 \hat{W}\hat{U}_0(\xi_1-\xi_2) \hat{W}\hat{U}_0(\xi_2)
\nonumber\\
...&&\nonumber
\end{eqnarray}
and where we defined $\hat{U}_0(\xi)\equiv
\exp\left[(\pi/2)(\hat{\sigma}_{+}-\hat{\sigma}_{-})\xi\right]$.

Using this approach we calculate the final states for various
initial states of the atom. We find that up to $O(\lambda^2)$ it is
enough to work in the  truncated bipartite basis that includes only
4 states:
$\left\{|g,0\rangle,|g,1\rangle,|e,0\rangle,|e,1\rangle\right\}$.
States of the form $|\sigma,2\rangle$ do appear in the expansion.
However, in our calculations of the entanglement and gate error they
do not contribute  up to and including  $O(\lambda^2)$ and are
therefore omitted. We thus end up with a truncated basis that,
following the perturbative assumption, includes only a single
photonic excitation in the pulse mode, as argued in \cite{SD-PRA69}.
The resulting final state $|\psi\rangle$ is not normalized, we
therefore expand the normalization factor
$n(\lambda)=1/\sqrt{\langle\psi|\psi\rangle}$ to second order in
$\lambda$ and use it to normalize the state.

\subsubsection{Entanglement calculation}
We measure the entanglement of the bipartite state using the
\emph{tangle} monotone which is related to the square of the
concurrence \cite{osborne,wooters} and compare our results for SL
pulses with those obtained in \cite{SD-PRA69} for CL pulses.

For a pure bipartite state $|\psi\rangle$ of two subsystems A and B
and where one of the subsystems is a two-level system, the tangle
takes the form \cite{rungta}
\begin{equation}
T \left (|\psi\rangle\langle\psi| \right )=2 \left (1-\mathrm{Tr}_A
\left [ \left (\mathrm{Tr}_B \left [|\psi\rangle\langle\psi| \right]
\right)^2 \right ] \right )\label{T}.
\end{equation}
We calculate the atom-pulse tangle for various atomic initial states
up to $O(\lambda^2)$, by inserting the normalized final state,
calculated using Eq. (\ref{evol}), into
Eq. (\ref{T}).\\
For atomic initial states $|g\rangle$ and $|e\rangle$  we get
\begin{equation}
T\approx
\left[1+\left(\frac{s}{c}\right)^2-2\left(\frac{s}{c}\right)\cos(2\phi)\right]c^2\kappa\tau.
 \label{92}
\end{equation}
For a CL pulse, i.e. $s=0$ and $c=1$, the tangle is given by
$T\approx\kappa\tau$, which is the same as calculated in
\cite{SD-PRA69}. For non-zero squeezing this result is rather
strongly modified and in particular \emph{the tangle depends on the
squeezing phase $\phi$}. As an example when $e^{r}\gg 1$ and
therefore $s/c$ is close to unity, we obtain $T\approx
2c^2\kappa\tau [1-\cos(2\phi)] $. This expression has a typical interference pattern which will be discussed in section III B4. \\

\begin{figure}
\begin{center}
\includegraphics[scale=0.67]{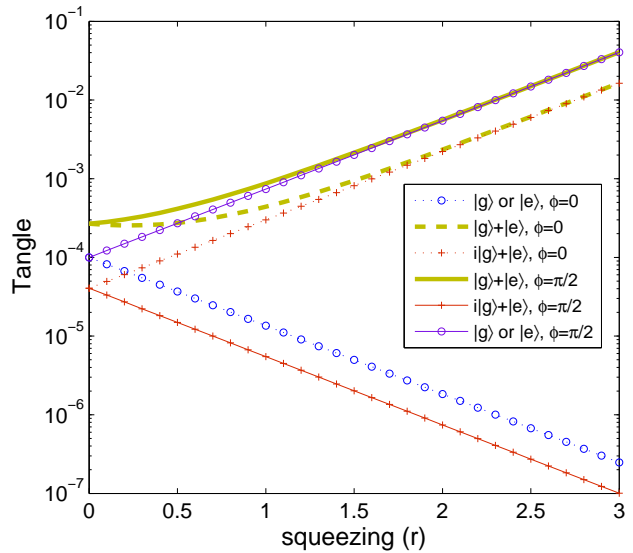}
\caption{\small{The tangle, Eq. (\ref{T}), calculated for the atom
and the squeezed light pulse mode bipartite system vs the squeezing strength. The
pulse duration is $\tau=0.1\Gamma^{-1}$
 and the coupling to the rest of the EM modes is neglected. The decay rate into paraxial modes is taken to be $\kappa=\Gamma/1000$. Results are shown for four atomic
  initial states and for the squeezing phases $\phi=0,\pi/2$, corresponding to amplitude and phase squeezing respectively.
 Compared with the coherent light case $(r=0)$, one obtains a
 squeezing-phase-dependent suppression and enhancement of the entanglement as
squeezing gets larger.}} \label{fig1_T}
\end{center}
\end{figure}

Let us now consider two different squeezing phases, $\phi=0$ and
$\phi=\pi/2$, which for $\theta=0$ correspond to amplitude and phase
squeezing respectively. We obtain from (\ref{92})
\begin{equation}
T\approx \kappa\tau e^{\mp 2r}  ,
 \label{93}
\end{equation}
where the minus (plus) sign is for $\phi=0$ ($\phi=\pi/2$).
When compared with the CL case, in the $\phi=\pi/2$ case we get
 \emph{enhancement} whereas in the $\phi=0$ case we get
\emph{suppression} of entanglement. This result is plotted in Fig.
\ref{fig1_T}, where we set $\tau=0.1\Gamma^{-1}$ and
$\kappa=\Gamma/1000$. This estimate for $\kappa$ results from
assuming a Gaussian beam focused on the atom with a lens of
numerical aperture of about 0.l, consistent with the paraxial
approximation \cite{VE-K,SD-PRA69}. With these numbers, the
assumption $\lambda\ll1$ is valid for $r\leq 3$. Current
state-of-the-art light squeezing experiments do not exceed
$r=1.16$ (see Sec. IV), therefore $\lambda\ll1$ is a very good assumption.

We repeat the tangle calculation for initial superposition states of
the form
$\frac{1}{\sqrt{2}}\left(e^{i\theta_A}|g\rangle+|e\rangle\right)$.
In a Bloch sphere representation this state is represented by a
vector lying in the equatorial $xy$ plane with an angle $\theta_A$
from the positive $x$ direction. The tangle we get is
\begin{equation}
T\approx
\frac{e^{-i2\theta_A}}{4\pi^2}\left[\pi\left(1+e^{i2\theta_A}\right)e^{\mp
r}+4e^{i\theta_A}e^{\pm r}\right]^2\kappa\tau,
 \label{94}
\end{equation}
where the upper (lower) sign refers to $\phi=0$ ($\phi=\pi/2$). This
expression is plotted in Fig. 1 for the two initial states
$\theta_A=0$ and $\theta_A=\pi/2$. For each initial state, a
squeezing-phase-dependent enhancement and suppression compared with
the CL case are observed.

As could be expected from atom-SL interaction, we observe dependence
on the relative phase between the squeezing phase, $\phi$, and the
phase of the average of the pulse, $\theta$  (which is chosen here
to be $0$). Furthermore, the tangle indeed depends on the atomic
initial state. For fixed pulse phase, the change in entanglement for
initial states that lie in the Bloch sphere equatorial $xy$ plane
depends on the difference between the squeezing phase and
$\theta_A$.

\subsubsection{Error probability calculation}
The error probability of a quantum operation is just one minus the
probability of the atomic qubit to be in the target state
$|\eta\rangle$  of the operation after the operation is performed. A
classical light pulse can be chosen to transform an atom from the
initial  to a prescribed final state $|\eta\rangle$. When operating
with the quantized pulse,  the operation is subject to quantum
fluctuations. The fidelity $F$ is the probability of the atom to be
in $|\eta\rangle$ after the
 \emph{quantized} field has operated on it, $F=\langle\eta|\hat{\rho}_A|\eta\rangle$,
 where $\hat{\rho}_A$ is the density matrix that describes the, generally
 mixed,
 final state of the two-level atomic system.
 The error probability is then,
\begin{equation}
P_{error}=1-F=1-\langle\eta|\hat{\rho}_A|\eta\rangle. \label {Per}
\end{equation}
Since in the absence of the reservoir the sole error is that due to
atom-pulse entanglement, we were able to find a simple analytic
relation between the tangle and the error probability calculated up
to $O(\lambda^2)$, $T\approx 4P_{error}$. This linear relation
implies that the results for $P_{error}$ follow similar trends to
those for the tangle and are \emph{squeezing phase} and initial
atomic state dependent (Fig. 2). For quantum information processing
purposes the relevant error is that averaged over all possible
atomic initial states, $<P_{error}>$. The averaged error is
calculated analytically using the fidelity of 6 different initial
atomic states \cite{avF}
\begin{equation}
<P_{error}>\approx\left(0.0675e^{\pm 2r}+0.1665e^{\mp
2r}\right)\kappa\tau,
\label{Pav}
\end{equation}
where as before the upper (lower) sign is for $\phi=0$
($\phi=\pi/2$). This result is plotted as a function of squeezing in
Fig. 3. As can be seen the averaged error, $<P_{error}>$, depends on
the squeezing phase $\phi$. For most values of the squeezing
parameter $r$ the average error is larger than in the CL case
($r=0$). This implies that on average, more entanglement is
generated between the atom and the pulse in SL than in CL case.
Still one observes that  for the region $0<r<0.45$, and $\phi=0$
(amplitude squeezing), the average error drops relative to the error
in the CL case. Though practically small, in this range of the
squeezing parameter the SL operation has \emph{lower quantum limit}
to the operation error than with the CL pulse due to reduced
intensity noise.

\begin{figure}
\begin{center}
\includegraphics[scale=0.67]{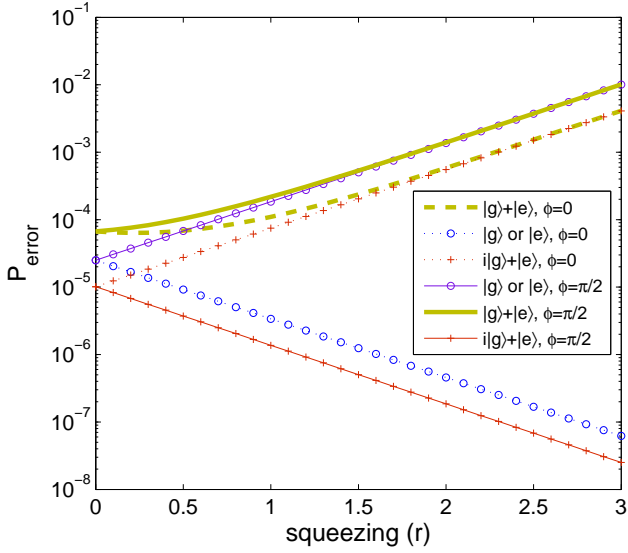}
\caption{\small{The error probability of an operation with a squeezed light $\pi$ pulse on an atom vs squeezing strength. The pulse
duration is  $\tau=0.1\Gamma^{-1}$ and the rest of the EM modes are disregarded. As in Fig. 1, $\kappa=\Gamma/1000$ is taken.
 The inset shows the atomic initial states and the squeezing phase. Here $\phi=0$ ($\pi/2$) corresponds to reduced amplitude (phase) fluctuations. The
 dependence on the squeezing phase follows the results for the
entanglement.  Lower errors than for the coherent light operation
($r=0$) can be achieved.}} \label{fig2a}
\end{center}
\end{figure}

\begin{figure}
\begin{center}
\includegraphics[scale=0.67]{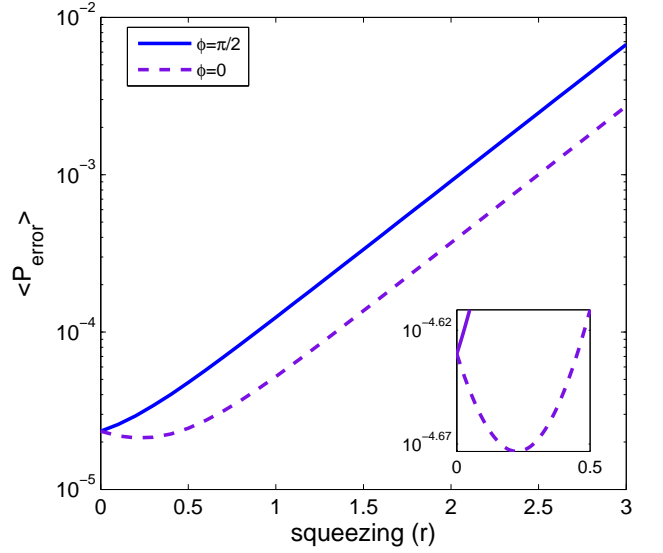}
\caption{\small{Same as Fig. 2, but for the average error probability.
 The average is over all possible atomic initial states. For most cases, the average error with squeezed light increases relative to coherent light ($r=0$). For the squeezing phase
 $\phi=0$ (amplitude squeezing)
 and squeezing strength $0<r<0.45$ a very small decrease in the average error
 is observed, as shown in the figure inset.}} \label{fig2b}
\end{center}
\end{figure}

\subsubsection{Discussion of the results}

\emph{4a. Quantum interference}\\
Consider the interaction $\hat{W}$ between the atom and the
fluctuating part of the light pulse shown in Eq. (\ref{F}). In
addition to the Jaynes-Cummings like atom-photon interaction term
$\hat{W}_1\equiv i(\hat{a}\hat{\sigma}_{+}-h.c.)$, present also in
the CL case, it contains a squeezing-dependent term $\hat{W}_2\equiv
i\left[(s/c)e^{-2\varphi}\hat{\sigma}_{+}\hat{a}^{\dag}-h.c.\right]$.
Atomic and light mode excitations can be generated via both
$\hat{W}_1$ and $\hat{W}_2$. These two paths interfere quantum
mechanically resulting a dependence of atom-light entanglement and
error probability on both squeezing and atomic phases.

Starting from $|g,0\rangle$ we examine the bipartite state evolution
under the first two operators in  Eq. (\ref{evol}).  The classical
field term, $\hat{U}^{(0)}$, is  evolving the state to $|e,0\rangle$
via superpositions of $|g,0\rangle$ and $|e,0\rangle$.
Simultaneously the state evolves under the term $\hat{W}$ in
$\hat{U}^{(1)}$. The $\hat{W}_1$ part of it  couples only to
$|e,0\rangle$ evolving it to $|g,1\rangle$. The $\hat{W}_2$ part on
the other hand couples only to $|g,0\rangle$ and transform it to
$|e,1\rangle$. Subsequent evolution under $\hat{U}_0$  forms
superpositions of $|e,1\rangle$ and $|g,1\rangle$, the amplitudes of
which are coherent sums (and therefore show interference) of
evolution amplitudes under both $\hat{W_1}$ and $\hat{W_2}$.
Quantitatively,
 \begin{eqnarray}
 &&|\psi(\tau)\rangle\approx\left[\hat{U}^{(0)}+\lambda \hat{U}^{(1)}\right]|g,0\rangle= \nonumber \\
 &&|e,0\rangle +\lambda \int_{0}^{1}d\xi \hat{U}_0(1-\xi)(\hat{W}_1 +\hat{W}_2)\hat{U}_0(\xi)|g,0\rangle = \nonumber \\
&&|e,0\rangle -\lambda\int_{0}^{1}d\xi \hat{U}_0(1-\xi)\sin\left(\frac{\pi}{2}\xi\right)|g,1\rangle-\nonumber\\
 &&\lambda\int_{0}^{1}d\xi \hat{U}_0(1-\xi)
 e^{-i2\phi}\frac{s}{c}\cos\left(\frac{\pi}{2}\xi\right)|e,1\rangle=\nonumber\\
 &&|e,0\rangle-\lambda
\frac{1}{\pi}\left(e^{-i2\phi}\frac{s}{c}+1\right)|e,1\rangle
 +  \lambda \frac{1}{2}\left(e^{-i2\phi}\frac{s}{c}-1 \right)|g,1\rangle, \nonumber\\
 \label{g_evol}
  \end{eqnarray}
where  $\hat{U}_0(\xi)|g/e,0\rangle=\cos\left(\frac{\pi}{2}\xi\right)|g/e,0\rangle\pm\sin\left(\frac{\pi}{2}\xi\right)|e/g,0\rangle$.\\
The interference between evolution under $\hat{W}_1$ and
$\hat{W}_2$ is clearly seen in the amplitudes of $|e,1\rangle$ and
$|g,1\rangle$. For large squeezing, i.e. $s\cong c$, and
$\phi=0$, we get
$|e\rangle\otimes\left(|0\rangle-\frac{2}{\pi}\lambda|1\rangle\right)$
which contains no entanglement.  For $\phi=\pi/2$ we get
$|e,0\rangle+\lambda|g,1\rangle$ which is entangled with
entanglement strength on the order of the interaction strength
$\lambda$. This result is similar to that obtained for the tangle in
III B2. For other atomic initial states similar calculations up to
second order in $\lambda$ were performed leading to the results
shown in III B2 and III B3.
\\\\

\emph{4b. Optical Bloch equation (OBE) with a noisy field}\\
An intuitive understanding of the error results from Sec. III B can
be obtained from the Bloch sphere picture. To this end we derive the
OBE for a one mode SL field, starting from the  Hamiltonian
(\ref{H_S1}), $\hat{H}=i\hbar
g\left[(\hat{a}+\alpha)\hat{\sigma}_{+}-(\hat{a}^{\dag}+\alpha^{\ast})\hat{\sigma}_{-}\right]$.
As usual for SL, we define the quadrature operators
$\hat{X}_{1,2}=\hat{\delta X}_{1,2}+x_{1,2}$ with
 \begin{eqnarray}
\hat{a}&\equiv&\frac{1}{2}\left(\hat{\delta X}_1+i\hat{\delta X}_2\right)\nonumber\\
\alpha&\equiv&\frac{1}{2}\left(x_1+ix_2\right).
\label{quad}
\end{eqnarray}
 Here $\hat{\delta X}_{1,2}$ are hermitian operators with zero average and $x_{1,2}$ are real numbers representing the averages of the quadratures. We then define the vectors
\begin{eqnarray}
\vec{x}&\equiv&\left(x_2,x_1,0\right)\nonumber\\
\hat{\vec{\delta X}}&\equiv&\left(\hat{\delta X}_2,\hat{\delta X}_1,0\right),
\label{rabi}
\end{eqnarray}
so that the Hamiltonian (\ref{H_S1}) becomes
\begin{equation}
\hat{H}=- \hbar\frac{g}{2}\left(\vec{x}+\hat{\vec{\delta X}}\right)\cdot\hat{\vec{\sigma}},
\end{equation}
with $\hat{\vec{\sigma}}=(\hat{\sigma}_x,\hat{\sigma}_y,\hat{\sigma}_z)$. The resulting Heisenberg equation for $\hat{\vec{\sigma}}(t)$ is
\begin{equation}
\frac{d\hat{\vec{\sigma}}}{dt}=- \frac{g}{2}\vec{x}\times\hat{\vec{\sigma}}- \frac{g}{2}\hat{\vec{\delta X}}\times\hat{\vec{\sigma}},
\label{sigma_t}
\end{equation}
with all the operators written in the Heisenberg picture. Rescaling the time to $\bar{t}=t/\tau$ Eq. (\ref{sigma_t}) becomes
\begin{equation}
\frac{d\hat{\vec{\sigma}}}{d\bar{t}}=- \vec{\Omega}\times\hat{\vec{\sigma}}- \frac{1}{2}\zeta \hat{\vec{\delta X}}\times\hat{\vec{\sigma}},
\label{sigma_tbar}
\end{equation}
where $\vec{\Omega}=(g\tau \vec{x}/2)$ is the dimensionless Rabi
vector and $\zeta=g\tau$. Recall that for a $\pi$ pulse
$|\alpha|=(\pi/2g\tau)$ and therefore $|\vec{\Omega}|=O(1)$. Our
calculation also assumes $\lambda=\cosh(r)\zeta\ll1$, i.e. even for
matrix elements of $\hat{\vec{\delta X}}$ which are of order
$\cosh(r)$, $\zeta\hat{\vec{\delta X}}$ is still very small. We can
then expand the operators
$\hat{\vec{\sigma}}(\bar{t})=\hat{\vec{\sigma}}_0(\bar{t})+\zeta\hat{\vec{\sigma}}_1(\bar{t})+...$
and $\hat{\vec{\delta X}}(\bar{t})=\hat{\vec{\delta
X}}_0(\bar{t})+\zeta\hat{\vec{\delta X}}_1(\bar{t})+...$, to obtain
to first order in $\zeta$,
\begin{equation}
\frac{d\hat{\vec{\sigma}}}{d\bar{t}}\approx- \vec{\Omega}\times\hat{\vec{\sigma}}- \frac{1}{2}\zeta \hat{\vec{\delta X}}_0\times\hat{\vec{\sigma}}_0.
\label{sigma_perturb}
\end{equation}
The last term includes the cross product of the Heisenberg picture operators in the absence of atom-field interaction,
 so that $\hat{\vec{\sigma}}_0(\bar{t})$ is simply the solution of the OBE with the classical field $\vec{\Omega}$
 and describes evolution in the atomic part of the Hilbert space, and $\hat{\vec{\delta X}}_0=\hat{\vec{\delta X}}(\bar{t}=0)$ is in the photonic part of the Hilbert space.

The average over atomic degree of freedom can be defined on any
operator $\hat{O}$ to be $\langle
\hat{O}\rangle_A\equiv\mathrm{tr_A}[\hat{\rho_A} \hat{O}]$, where
$A$ denotes the atom system and $\hat{\rho_A}$ is the reduced atomic
density operator. Averaging equation (\ref{sigma_perturb}) this way
we get an OBE with a noisy field,
 \begin{equation}
\frac{d\langle\hat{\vec{\sigma}}\rangle_A}{d\bar{t}}\approx-
\vec{\Omega}\times\langle\hat{\vec{\sigma}}\rangle_A-\frac{1}{2}\zeta
\hat{\vec{\delta X}}_0\times\langle\hat{\vec{\sigma}}_0\rangle_A
\label{OBE}.
\end{equation}
Note that $\langle\hat{\vec{\sigma}}\rangle_A(\bar{t})$ and
$\hat{\vec{\delta X}}_0$ are operators in the photonic space and are
therefore fluctuating, whereas
$\langle\hat{\vec{\sigma}}_0\rangle_A(\bar{t})$ is just the c-number
Bloch vector solution of an OBE driven by the average field
$\vec{\Omega}$ and it in fact equals to the average of
$\hat{\vec{\sigma}}(\bar{t})$ over both atomic and photonic degrees
of freedom. The term with $\hat{\vec{\delta X}}_0$ is the one driven
by the fluctuations of the field, which in our case are quadrature
dependent SL fluctuations. For $\phi=0$, the uncertainties of the
quadratures are $\Delta \hat{\delta X}_1=e^{-r}\leq1$ and $\Delta
\hat{\delta X}_2=e^{r}\geq1$, i.e. $\hat{\delta X}_1$ has reduced
fluctuations and $\hat{\delta X}_2$ has increased fluctuations
relative to CL. For $\phi=\pi/2$ we get the opposite.

We can now present the field in the Bloch sphere as seen in Fig. 4.
Taking real $\alpha$ $(\theta=0)$ and squeezing phase $\phi=0$, the
Rabi vector is pointing to the $y$ direction $(x_2=0)$ and is
amplitude squeezed ($\Delta\hat{\delta X}_1<1$). Now consider an
initial ground state, i.e. $\langle\hat{\vec{\sigma}}\rangle_A$ is
pointing to the $-z$ direction. For a $\pi$ pulse operation, the
$xy$ phase of the field is not relevant, and therefore phase noise
of the field doesn't affect the operation. The only relevant noise
is the amplitude noise which for $\phi=0$ is reduced relative to CL
case. We thus expect to get a lower error in this operation as
compared with the CL operation. For $\phi=\pi/2$, the field has an
increased amplitude noise ($\Delta\hat{\delta X}_1>1$), which
results an increased error compared to the CL case. Now consider the
$y$ axis pointing initial state again for $\theta=0$. Since the Rabi
vector is also pointing to the $y$ direction the average "moment"
obtained from the cross product is zero. Here only the $\hat{\delta
X}_2$ "phase" noise, in the $x$ direction, affects the qubit state.
When this noise is increased ($\phi=0$) the error is larger than in
the CL case, and when it is reduced ($\phi=\pi/2$) the error gets
lower. For an initial state pointing to the $x$ direction, one can
see that in general, both $\hat{\delta X}_2$ and $\hat{\delta X}_1$
affect the operation, and therefore squeezing in any quadrature
typically increases the operation error.

\begin{figure}
\centering
\includegraphics[scale=0.5]{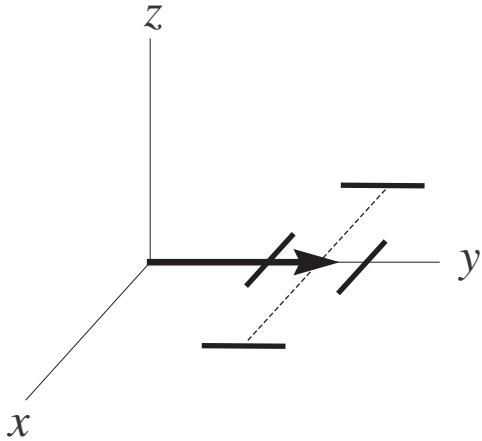}
\caption{\small{The driving field in the optical Bloch equation
representation. The solid arrow is the Rabi vector $\vec{\Omega}$
proportional to $\vec{x}$ from Eq. (\ref{rabi}) for $x_2$=0
($\theta=0$). The error bars are the variances of the field
fluctuations $\Delta\hat{\delta X}_2$ and $\Delta\hat{\delta X}_1$
in the $x$ and $y$ directions respectively. Here $\phi=0$ is taken,
i.e. $\Delta\hat{\delta X}_1<\Delta\hat{\delta X}_2.$}}\label{fig4}
\end{figure}

\subsection{Entanglement and error probability calculations in the presence of the reservoir}
We now wish to examine how accounting for the spontaneous emission
into the reservoir of empty modes will modify the above results. To
this end the full master equation, Eq. (\ref{master}), has to be
solved. We again assume $\lambda\ll1$ and work in the truncated
basis
$\left\{|g,0\rangle,|g,1\rangle,|e,0\rangle,|e,1\rangle\right\}$. We
solve Eq. (\ref{master}) numerically in this basis using
$\kappa=\Gamma/1000$ and calculate the density matrix
$\hat{\rho}(\tau)$ of the bipartite final state. We then use
$\hat{\rho}(\tau)$ to evaluate the tangle and the error probability.

\subsubsection{Tangle calculation}
\begin{figure}
\centering
\includegraphics[scale=0.6]{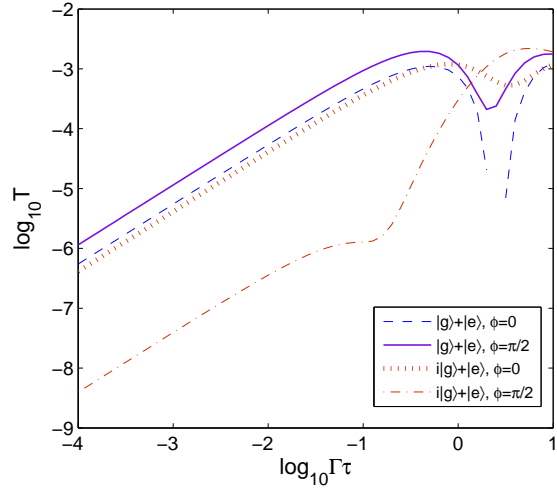}
\caption{\small{The tangle generated between the atom and the
squeezed light pulse mode as a function of the pulse duration scaled
to atom free space decay time $\Gamma^{-1}$. The squeezing of the
pulse mode is taken to be $r=1.16$,
 which corresponds to $10\log_{10}(e^{2r})=10\mathrm{dB}$ quadrature noise reduction. As in Fig. 1, $\kappa=\Gamma/1000$ is taken. The
 reservoir of the free space EM modes is included. Results
 for two
  atomic initial states are plotted. The effect of the reservoir is seen in the deviation from the initial linear part of the plots. This deviation begins as expected at times slightly shorter than $\tau\sim\Gamma^{-1}$, when entanglement is enhanced
 by SL (all cases here, apart from $i|g\rangle+|e\rangle$, $\phi=\pi/2$). When
 entanglement is suppressed by SL, the effect of the reservoir is seen already at much smaller $\tau$.}} \label{Tir}
\end{figure}

Within the truncated basis, the  pulse mode is also a two-level
system. We follow \cite{osborne, wooters} where a procedure for the
tangle calculation of a two-qubit mixed state is given. Figure
\ref{Tir} presents the tangle as a function of the pulse duration
$\tau$. We assume squeezing of $r=1.16$, which corresponds to
$10\log_{10}(e^{2r})=10\mathrm{dB}$ pulse mode quadrature noise
reduction. The interaction with the reservoir changes atom-pulse
entanglement when $\Gamma\tau\gtrsim 1$ and should generally
decrease it for $\Gamma\tau\gg 1$. Whenever the tangle is not
suppressed by the SL operation, pulses shorter than $0.1\Gamma^{-1}$
are well accounted by neglecting the reservoir, i.e. extending the
initial linear part of the plots. When SL operation suppresses the
entanglement, the interaction with the reservoir becomes more
dominant, and the effect of the reservoir can only be neglected for
pulses shorter than $0.01\Gamma^{-1}$.

\subsubsection{Error probability calculation}

\begin{figure}
\centering
\includegraphics[scale=0.45]{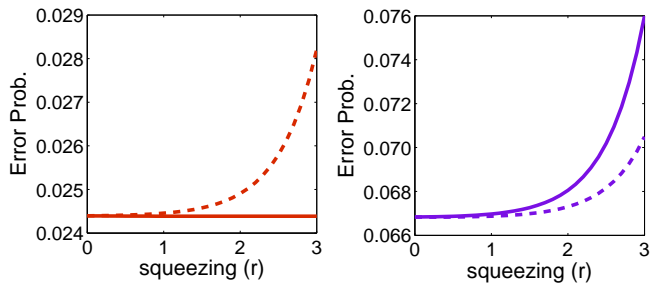}
\caption{\small{The error probability of an operation with a
squeezed light $\pi$ pulse on an atom vs squeezing strength. The
pulse duration is  $\tau=0.1\Gamma^{-1}$ and
 atomic decay into all the reservoir modes is considered. As in Fig. 1, $\kappa=\Gamma/1000$ is taken. The left plot
  is the calculation for $i|g\rangle+|e\rangle$
   atom initial state and the right one is for $|g\rangle+|e\rangle$.
    The dashed lines are for $\phi=0$ and the solid lines for $\phi=\pi/2$. }} \label{Pir}
\end{figure}

The error probability is calculated using Eq. (\ref{Per}). Figure
\ref{Pir} shows the results for  $\tau=0.1\Gamma^{-1}$ pulse
duration. Let us now compare between Fig. \ref{Pir} and Fig. 2,
where the reservoir has been neglected. We focus on the case of the
initial state $\frac{1}{\sqrt{2}}\left(i|g\rangle+|e\rangle\right)$.
For $r=0$, i.e. a CL pulse, the error of about $0.0244$ seen in the
left hand part of Fig. \ref{Pir} is actually due to spontaneous
emission to all the EM modes, as can be understood from the Mollow
picture (see Ref. \cite{itano}). The corresponding error in Fig. 2
is only $10^{-5}$. This suggests that owing to its small solid angle
($\kappa=\Gamma/1000$), the error due to coupling to the laser mode
is overwhelmed by that due to coupling to the reservoir modes. When
squeezing is introduced ($r>0$), for $\phi=\pi/2$, the atom-pulse
entanglement is suppressed as  seen in Fig. 2.  This suppression is
hardly noticeable when the reservoir modes are included as one can
see from Fig. \ref{Pir}. Numerically, for $r=3$, we find that the
error decrease is $0.95\times10^{-5}$. This matches what could be
expected from Fig. 2 and demonstrates that the reservoir is  the
major cause of error in this case. On the other hand, for $\phi=0$,
an increase of the error seen in Fig. 2 persists also in the
presence of the reservoir, as can be seen in  Fig. \ref{Pir}. For
very large squeezing this increase may become of the same order of
magnitude as the error due to the spontaneous emission to all the EM
modes. E.g. for $r=3$ the increase seen in both Figs. 2 and
\ref{Pir} is $\sim 4\times 10^{-3}$, about $16.3\%$ of the $0.0244$
error induced by spontaneous emission.

\section{Discussion}\label{section4}

The results of Sec. III show that the entanglement between the pulse mode and
 the atom is indeed modified in a squeezing-phase-dependent way. However, the use
 of squeezed light would not reduce the average
error compared with the CL case by a practicably useful amount.
Despite its inapplicability for Quantum computing, the entanglement
between SL and a single atom is of scientific interest, and would be
interesting if observed experimentally.

 Contrary to CL, where
atom-light entanglement is probably too small to be detected, SL
with a large squeezing parameter $r$ and a right choice of phases,
will eventually enhance SL-atom entanglement to experimentally
detectable values. A direct measurement of entanglement would be
through complete state tomography of the combined SL-atom state. The
tangle would then be calculated from the reconstructed density
matrix. The requirements for this experiment would be a squeezed
light source with a sufficiently large squeezing parameter
\cite{squeezing-bandwidth}, atomic state detection and
photodetectors with sufficient efficiency. Alternatively, ignoring
the light mode, the error in the final atomic state can be measured
for various parameters, such as different squeezing phases, thus
relaxing the need for an efficient photodetector.

Examples for squeezed light sources include four wave mixing in a
nonlinear medium, optical parametric amplifiers, second harmonic
generation and amplitude squeezed light from diode lasers among
others \cite{Bachor}. All these sources can be pulsed, e.g. via
electro-optic modulators. The initial light state we assumed for the
control field was a single-mode squeezed state. Parametric down
conversion (PDC) of short pulses could be considered as one example
of a realization for such a light source. In \cite{boyd, wasilewski}
the multimode pulsed SL produced by PDC is analyzed and described by
a discrete set of squeezed modes. In fact, for some conditions of
phase matching and pump pulse duration, only one of the modes of the
discrete set is squeezed, resulting in a pulse in a single-mode
squeezed state.

Even the most advanced single-mode squeezed light sources to-date
cannot reach quadrature noise suppression greater than $10$dB
($r=1.16$) \cite{Bachor}. This means that the maximal error increase
and tangle expected and therefore the minimal required detection
error, will be of the order of a few $10^{-4}$ (see Fig. 1 and 2).
Current state-of-the-art atomic state detection has an error which
is of similar magnitude \cite{Langer, Oxford}. However the
efficiency of even the most advanced photodetectors is far from
satisfying such low error rates. It therefore seems that a direct
tangle measurement is currently unfeasible, however, an indirect
indication for entanglement, detected through gate error
enhancement, is within experimental reach.

One important limitation for observing the error modification
effects in free space is the spacial overlap between the atom and
the paraxial pulse mode. Similar limitations make up a major
difficulty in observing the effect of the inhibition of the atom
coherences decay in a squeezed vacuum reservoir, found by Gardiner
\cite{gar86}. In this context, other schemes, involving
atom-squeezed vacuum interaction in a cavity, were considered
\cite{ParZollerCar}. Recent theoretical analysis and experiment
however have shown a possibility of strong coupling of a laser mode
to a single atom in free space \cite{vanEnkKimble2000, Kurtsiefer,
Agio}. The use of such beams, where the overlap between the incoming
beam and the atomic dipole radiation pattern is large ($\kappa\sim
\Gamma$), may greatly enhance the effects discussed above.

In many experiments, a pair of ground, e.g. hyperfine, states of an
atomic system is used as a qubit. Although the energy separation
between ground states is typically in the radio-frequency range,
quantum gates are often driven with a laser beam pair, off-resonance
with a third optically excited level, in a Raman configuration. The
limitations to hyperfine qubit coherence and gate fidelity due to
the quantum nature of the Raman laser fields was investigated both
experimentally and theoretically \cite{Ozeri 2005, Ozeri 2007}. In
this case too, it is interesting to ask whether the use of, either
independently or relatively, squeezed light fields would influence
the gate error. An experiment, similar to that described above, but
with a ground state qubit could benefit from the ground state qubit
intrinsic longevity. In fact, a gate averaged error of few
$10^{-3}$, dominated by laser classical noise and photon scattering,
was recently measured on a single trapped-ion hyperfine qubit and a
Raman laser gate \cite{Knill 2008}.
\\

To conclude, we have calculated the entanglement between a short
pulse of resonant squeezed light and a two-level atom in free space
during a qubit $i\hat{\sigma}_y$ gate (a $\pi$ pulse) and the
resulting gate error. Several differences arise when comparing to
coherent control with CL. As could be expected for squeezed light,
three phases determine the error magnitude - the squeezing phase,
the light phase and the phase of the initial qubit superposition. In
comparison with the coherent light case, the error and the
entanglement can be either enhanced or suppressed depending on the
above phase relations. These entanglement enhancement and
suppression effects naturally become stronger with squeezing and can
get to a few orders of magnitude. For quantum information processing
purposes the relevant error is that averaged over all possible
initial states. We have shown that although in principal the average
error can be reduced by using SL within a specific range of
parameters, the resulting minimum error however would be of the same
order of magnitude as that with CL. In fact, even when not averaged,
but when taking into account the free space environment, the error
is still mostly dominated by the atomic decay to the EM reservoir
owing to a very small atom-pulse spacial overlap, so that any
reduction is not likely do be detected experimentally. On the other
hand, the effect of enhanced atom-pulse entanglement is of
scientific interest and might even prove useful as a resource in
some quantum communication schemes. A measurement of the increased
error would be an indirect indication for the enhancement of
entanglement by squeezing.

\section*{ACKNOWLEDGEMENTS}
We appreciate useful discussions with Ilya Averbukh, Eran Ginossar
and Barak Dayan. RO acknowledges support from the ISF Morasha
program, the Chais family Fellow program and the Minerva foundation.

\appendix*
\section{Quantized Paraxial Approximation}
Here we would like to derive the model Hamiltonian of the atom-pulse
mode interaction in free space seen in Eq. (\ref{H_I}). We start
from the usual atom-field Hamiltonian in the dipole approximation
\begin{eqnarray}
\hat{H}&=&\hat{H}_F+\hat{H}_A+\hat{H}_{AF} \nonumber \\
\hat{H}_F&=&\sum_{\vec{k} \mu} \hbar\omega_k
\hat{a}^\dagger_{\vec{k} \mu} \hat{a}_{\vec{k} \mu}
\nonumber \\
\hat{H}_A&=&\frac{1}{2}\ \hbar\omega_{a}\hat{\sigma}_{z}\nonumber\\
\hat{H}_{AF}&=&\sum_{\vec{k}
\mu}i\sqrt{\frac{\hbar\omega_k}{2\epsilon_0}}d_{\hat{k}\mu} \left (
\hat{a}_{\vec{k} \mu}\frac{e^{i\vec{k}\cdot \vec{r}_a}}{\sqrt{V}} -
h.c. \right) \left ( \hat{\sigma}_{+} +
\hat{\sigma}_{-} \right ), \nonumber\\
\label{H_tot_k}
\end{eqnarray}
where $\vec{k}\mu$ are the indices of a Fourier mode and its
polarization respectively, and $\vec{r}_a$ is the position of the
atom. The dipole matrix element $\vec{d}$ is assumed to be real and
$d_{\hat{k}\mu}=\vec{\varepsilon}_{\hat{k}\mu}\cdot \vec{d}$, where
$\vec{\varepsilon}_{\hat{k}\mu}$ is the polarization vector with
polarization $\mu$ and is perpendicular to the propagation direction
$\vec{k}$. We first divide the sums on Fourier modes into the
paraxial subspace modes, $\Lambda_p$, and the nonparaxial subspace
modes. The paraxial subspace modes include the modes centered around
some carrier wave vector in the $z$ direction
$\vec{k}_0=k_0\vec{z}$:
\begin{eqnarray}
\Lambda_p&=&\{\Lambda_p^{x},\Lambda_p^{y},\Lambda_p^{z}\}\nonumber\\
\Lambda_p^{x}&=&(-k_x^{co},
k_x^{co})\nonumber\\\Lambda_p^{y}&=&(-k_y^{co},
k_y^{co})\nonumber\\\Lambda_p^{z}&=&(k_0-k_z^{co}, k_0+k_z^{co}),
\label{1}
\end{eqnarray}
where the bandwidths for (x,y,z) directions are narrow, i.e.
\begin{eqnarray}
&& k_i^{co}\ll k_o   \nonumber \\ && i=x,y,z.\label{2}
\end{eqnarray}
The nonparaxial modes are simply the rest of the modes $\vec{k} \mu$
not included in $\Lambda_p$ and we denote them with indices $\{q\}$.
The paraxial subspace $\Lambda_p$ can be approximately spanned by
the modes which are the solutions of the paraxial wave equations in
free space with carrier frequencies $\omega_k=(k_0+k)c$,
$TEM_{nm}^{k_0+k}$. These modes are denoted by the functions
$U_{nmk}(\vec{r})=u_{nmk}(\vec{r})e^{i(k_0+k)z}$. Their Fourier
coefficients are $U_{nmk}(\vec{k})$ so that their mode operators are
\begin{equation}
\hat{a}_{nmk,\mu}=\sum_{\vec{k}\in\Lambda_p}U_{nmk}^{\ast}(\vec{k})\hat{a}_{\vec{k}\mu}\Rightarrow
\hat{a}_{\vec{k}\mu}=\sum_{nmk}U_{nmk}(\vec{k})\hat{a}_{nmk,\mu}.\label{a_nmk}
\end{equation}
Note that the Fourier modes here, $\hat{a}_{\vec{k}\mu}$, are only
those included in the paraxial subspace. Using (\ref{a_nmk}) it is
easy to show within the paraxial approximation that
\begin{equation}
\sum_{\vec{k}\mu}\sqrt{\omega_k}d_{\hat{k}\mu}
\hat{a}_{\vec{k}\mu}\frac{e^{i\vec{k}\cdot
\vec{r_a}}}{\sqrt{V}}\approx \sqrt{\omega_0}
\sum_{nmk,\mu}d_{\mu}\hat{a}_{nmk,\mu}U_{nmk}(\vec{r}_a), \label{A1}
\end{equation}
where $\omega_0=k_0c$. We denote all the $\{nm\neq 00,k,\mu\}$ and
$\{nm=00,k,\mu=2\}$ modes by $\{i\}$, where $\{\mu\}=\{1,2\}$ are
the two polarizations. The rest of the modes ,$\{nm=00,k,\mu=1\}$,
have Gaussian transverse profile and are denoted $\{k\}$. We thus
get
\begin{eqnarray}
\hat{H}_{AF}&=&i \hbar \sum_q \left ( g_q \hat{a}_q -
g_q^{\ast}\hat{a}_q^{\dagger} \right) \left ( \hat{\sigma}_{+} +
\hat{\sigma}_{-} \right )  +  \nonumber \\&& i \hbar \sum_i \left (
g_i \hat{a}_i - g_i^{\ast}\hat{a}_i^{\dagger} \right) \left (
\hat{\sigma}_{+}
+ \hat{\sigma}_{-} \right ) + \nonumber \\
&& i\hbar \sum_k \left ( \sqrt{\frac{\omega_0}{2\epsilon_0 \hbar}}d
U_{00k}(\vec{r_a}) \hat{a}_k - h.c \right) \left ( \hat{\sigma}_{+}
+ \hat{\sigma}_{-} \right ), \nonumber \\ \label{H_tot_gk}
\end{eqnarray}
where $d=d_{\mu=1}$ and $g_i$, $g_q$ are the atom-mode dipole
couplings for the modes $\{i,q\}$. The dipole couplings for the
$\{k\}$ modes are seen explicitly in (\ref{H_tot_gk}),
\begin{equation}
g_k=\sqrt{\frac{\omega_0}{2\epsilon_0 \hbar}}d
U_{00k}(\vec{r}_a)=\sqrt{\frac{\omega_0}{2\epsilon_0 \hbar}}d
u_{00k}(\vec{r}_a)e^{i(k_0+k)z_a}. \label{g_k1}
\end{equation}
We note that $u_{00k}(\vec{r}_a)$ is just the overlap of the $k$
mode with the atom. This overlap includes some normalization of a
quantization length in the beam direction, $L$, which  should
eventually be taken to be infinity, and which is multiplied by some
effective area of the mode, $A$, evaluated at the atom's position,
i.e. $u_{00k}(\vec{r_a})=1/\sqrt{AL}$. Considering the narrow band
of $k$ values around $k_0$, $A$ is approximately independent of $k$.
The couplings $g_k$ are then written
\begin{equation}
g_k=d \sqrt{\frac{\omega_0}{2 \epsilon_0 \hbar A
}}\frac{1}{\sqrt{L}}e^{i(k_0+k)z_a}. \label{g_k}
\end{equation}
As for $\hat{H}_F$, since all the modes discussed here, $\{q,i,k\}$,
are eigenmodes of free space in the paraxial approximation, they
have well-defined single-particle energies $\hbar\omega_q$,
$\hbar\omega_i$ and $\hbar\omega_k$ respectively. The total
Hamiltonian from Eq. (\ref{H_tot_k}) thus becomes
\begin{eqnarray}
\hat{H}&=&\hat{H}_F+\hat{H}_A+\hat{H}_{AF} \nonumber\\
\hat{H}_F&=&\sum_q \hbar\omega_q \hat{a}^\dagger_{q} \hat{a}_{q} +
\sum_i \hbar\omega_i \hat{a}^\dagger_{i} \hat{a}_{i} +  \sum_{k}
\hbar\omega_{k} \hat{a}^\dagger_{k}
\hat{a}_{k}\nonumber \\
\hat{H}_A&=&\frac{1}{2}\ \hbar\omega_{a}\hat{\sigma}_{z}\nonumber\\
\hat{H}_{AF}&=&i \hbar \sum_{j=\{q,i,k\}} \left ( g_j \hat{a}_j -
g_j^{\ast}\hat{a}_j^{\dagger} \right) \left ( \hat{\sigma}_{+} +
\hat{\sigma}_{-} \right )     . \label{H_tot}
\end{eqnarray}
We would now like to describe a pulse mode with a Gaussian
transverse profile composed of Gaussian beams of different
frequencies around the carrier frequency $\omega_0$. The pulse mode
is then spanned by the $\{k\}$ modes. We can choose a new
orthonormal basis $\{f\}$ which spans the same subspace as that of
the $\{k\}$ basis and that includes a pulse mode $f_0$ as one of its
members. Both $\{k\}$ and $\{f\}$ modes share a similar Gaussian
$xy$ profile, we therefore focus only on their $z$ dependence. We
write the relation between the  $k$ and $f$ mode functions as
\begin{equation}
\Phi_{f}(z)=e^{ik_{0}z}\varphi_{f}(z)= e^{ik_{0}z}\sum_{k}
\varphi_{f}(k) \frac{e^{ikz}}{\sqrt{L}}, \label{phi_f}
\end{equation}
where $e^{i(k_0+k)z}/\sqrt{L}$ is a $k$ mode while $\Phi_{f}(z)$ is
an $f$ mode. An $f$ mode is thus defined by its envelope
$\varphi_{f}(z)$. In a similar way, the transformation for the field
operators is
\begin{equation}
\hat{a}_{k}= \sum_{f} \varphi_{f}(k) \hat{a}_{f} .\label{a_f}
\end{equation}
The pulse mode is then characterized by the operator
$\hat{a}_{f_0}$, whereas the Hilbert space of the two-level atom is
spanned by the Pauli spin matrices. The atom and the pulse mode make
for the relevant bipartite system. The rest of the EM modes,
$\hat{a}_{f\neq f_0}$, $\hat{a}_i$ and $\hat{a}_q$, interact with
the bipartite system via the atom and make up a reservoir.

We now focus on the part of the Hamiltonian that includes only the
atom and the Gaussian paraxial modes $\{k\}$, and their interaction
\begin{eqnarray}
\hat{H}^p&=&\sum_k \hbar\omega_k \hat{a}^\dagger_{k} \hat{a}_{k}
+\frac{1}{2}\hbar\omega_{a}\hat{\sigma}_{z} + \nonumber \\
&& i \hbar \sum_k\left ( g_k \hat{a}_k - h.c. \right) \left (
\hat{\sigma}_{+} + \hat{\sigma}_{-} \right ).\label{A5}
\end{eqnarray}
Moving to the interaction picture in the rotating wave
approximation, assuming no detuning ($\omega_a=\omega_0$) and using
equations (\ref{g_k}, \ref{phi_f}, \ref{a_f}), we get
\begin{equation}
\hat{H}_{I}^p=i \hbar \sum_{f}\left [\ g_f(t) \hat{a}_{f}
\hat{\sigma}_{+} - h.c. \right], \label{A6}
\end{equation}
with
\begin{equation}
g_f(t)=d\sqrt{\frac{\omega_{0}}{2\epsilon_{0}\hbar A}}
e^{ik_{0}z_a}\varphi_{f}(z_a-ct).
 \label{7}
\end{equation}
The system Hamiltonian of the relevant bipartite system comprised of
the atom and the pulse mode is then
\begin{equation}
\ \hat{H}_S\equiv \hat{H}_I^{f_0+atom}=i \hbar g_{f_0}(t) \left [\
\hat{a}_{f_0} \hat{\sigma}_{+} - h.c. \right], \label{A8}
\end{equation}
where we take the envelope of the pulse, $\varphi_{f_0}(z)$, to be
real and set $z_a=0$. Denoting all the rest the EM modes apart from
$f_0$ with indices $\{r\}$ we get the total Hamiltonian in the
interaction picture in Eq. (\ref{H_I}).


\begin{thebibliography}{}
\bibitem{squeezing-bandwidth} Noise reduction below the SQL was typpically
demonstrated at noise frequencies of above a few MHz, whereas at low
frequencies noise was dominated by classical laser noise. Here, for
the average error to remain small, noise reduction is necessary at
low frequencies as well.
\end{thebibliography}
\end{document}